\documentclass{knawproc}
\usepackage{epsfig}

\newcommand{\etal}{{\it et al.\,}}

\newcommand{\I}{{\small I}}
\newcommand{\II}{{\small II}}
\newcommand{\III}{{\small III}}
\newcommand{\IV}{{\small IV}}
\newcommand{\V}{{\small V}}

\begin{document}

\begin{opening}

\title{Spectroscopy of $z > 3$ Lyman--limit Galaxies}

\author{Hyron Spinrad$^\dag$\\
Arjun Dey$^\ddag$\\
Daniel Stern$^\dag$ \& Andrew Bunker$^\dag$}
\addresses{%
  $^\dag$Department of Astronomy, University of California, Berkeley \\
  $^\ddag$NOAO/KPNO, 915 N.~Cherry Ave., Tucson AZ~85726}

\runningtitle{$z>3$ Dropouts}
\runningauthor{Spinrad {\em et al.}}

\end{opening}


\begin{abstract}

We discuss the spectral character of Lyman--limit--selected,
star--forming galaxies at $z > 3$. The rest--frame UV spectra of these
faint galaxies may show Ly$\alpha$ in either absorption or emission,
probably depending upon their local ISM content and geometry.  Other UV
interstellar resonance absorption lines show considerable variation in
strength, likely related to differences in the galactic metal
abundances.

We present initial results on $B$--drop galaxies, generally at $z\sim
4$.  Our low--resolution spectrograms show no measurable flux below the
redshifted Lyman limit (912\,\AA).  Thus, it is likely that normal,
star--forming galaxies at early cosmic epochs did not significantly
contribute to the metagalactic ionizing radiation field.

\end{abstract}


\section{Introduction}

While not the ``main characters'' in the Academy colloquium, actively
star--forming galaxies are now routinely discovered and made available
for study through color--selection.  $U$-- and $B$--dropouts, galaxies
targeted for the redshifted Lyman limit spectral discontinuity at 912
\AA\ in the $U$-- and $B$--bands, are important in the early Universe
because of their ubiquity ({\em c.f.\ }Steidel {\em et al.\
}1996a,b). The $U$--drops are several percent of the total deep number
counts at $R \sim 24^{m}$ (corresponding to $B \sim 25^{m}$), with an
absolute surface density of $\sim 3$ galaxies per square arcminute on
the sky.  Moderate power radio sources are far less numerous, with
surface densities of $\sim 2\times 10^{-3}$ radio sources per square
arcminute at $S_{\rm 1.4~GHz} = 10$\,mJy.

In an informal collaboration with C.~Steidel and M.~Pettini, we have
pursued moderate resolution observations of the brightest ($R \sim
23^{m}$) $z\sim 3$ star--forming galaxies discovered by Steidel and
collaborators ({\em e.g.}, Steidel \& Hamilton 1992; Steidel {\em et
al.\ }1996a).  Typical sources are at $z\sim 3$ and require long
integrations at Keck with the LRIS spectrograph.  For the fainter and
(usually) more distant $B$--drop galaxies ($z\sim 4$), lower spectral
resolution is obtained. This work is part of an ongoing effort to study
`normal', {\em i.e.}, non--AGN, young galaxies at the earliest cosmic
epochs.

\begin{figure}[t]
\epsfxsize=5in
\epsffile{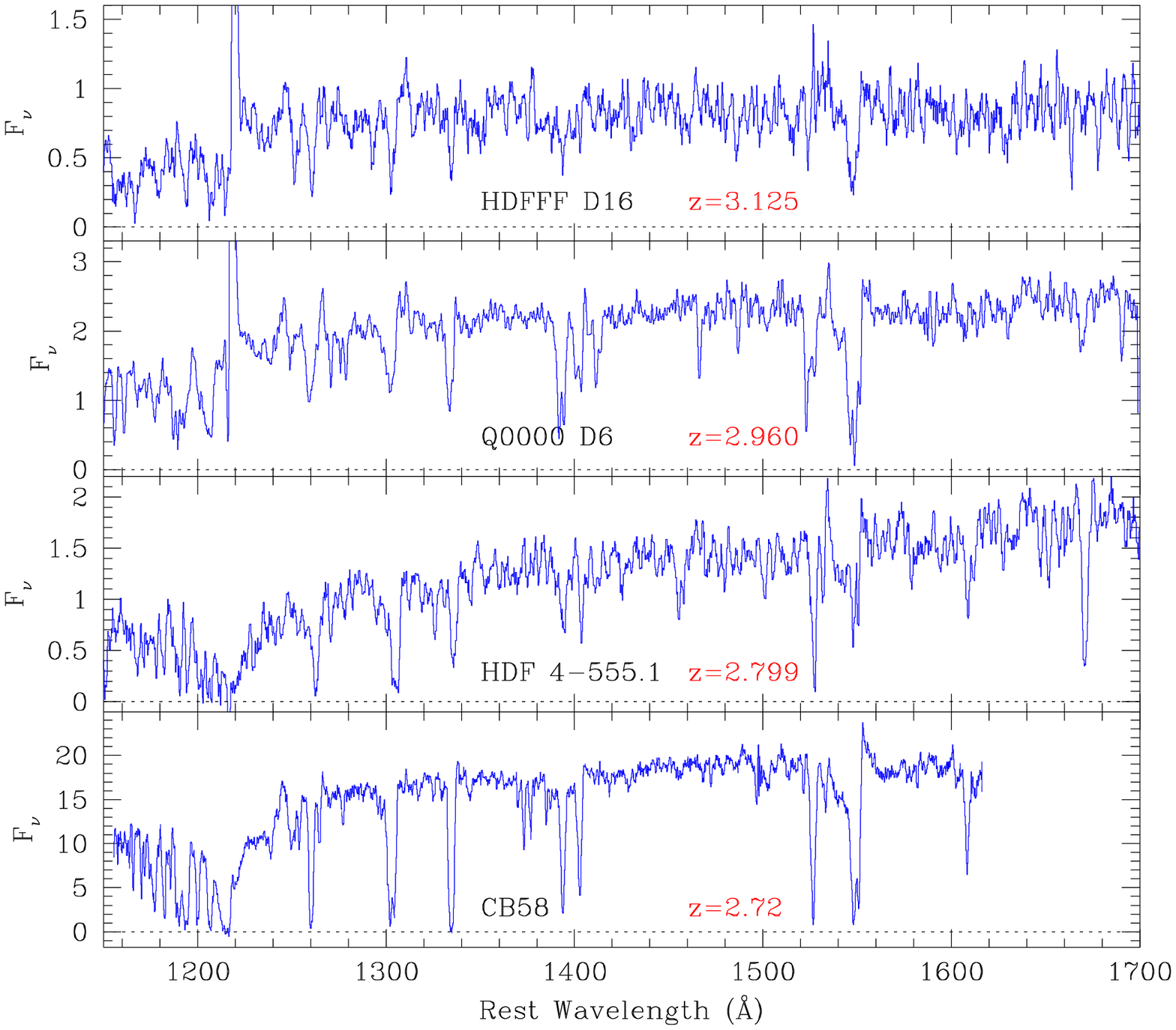}
\caption{Sequence of UV spectra of four Lyman--limit galaxies, arranged
from highest redshift to lowest redshift. Note the large range
in line strengths present.  In particular, Ly$\alpha$
($\lambda\,1216$\,\AA) is visible in both emission and absorption, while
primarily interstellar resonant lines (see Table~1) show considerable
variation in strength, possibly related to metallicity in these
young star--forming galaxies. The Ly$\alpha$ absorption galaxies also
appear redder in their continua at $\lambda>1300$\,\AA .}
\end{figure}

\section{Spectral Character of Lyman--Limit Galaxies}

The spectral character spanned by our current $U$--drop sample is
illustrated in Figure~1.  The galaxy spectra have been aligned in their
rest frames, so that the rather large range in emission and absorption
strengths is easily visible.  In particular, note the bimodal Ly$\alpha$
morphology; the variation in strength of the P--Cygni line profiles due
to O--star winds ({\em e.g.\
}C{\scriptsize~IV}\,$\lambda\lambda$\,1548,1551\,\AA ;
Si{\scriptsize~IV}\,$\lambda\lambda$\,1394,1403\,\AA); as well as the
variation of the primarily interstellar lines such as
Si{\scriptsize~II}\,$\lambda$\,1260\,\AA,
O{\scriptsize~I}\,$\lambda$\,1302\,\AA,
C{\scriptsize~II}\,$\lambda$\,1335\,\AA,
Fe{\scriptsize~II}\,$\lambda$\,1608\,\AA, and
Al{\scriptsize~II}\,$\lambda$\,1670\,\AA\ (see Table~1).

To be more specific, note that Ly$\alpha$ emission is strong and
slightly asymmetric with a broader red wing in HDF\,FF~D16 (from the HDF
flanking field) and Q\,0000~D6 (from a quasar field, Steidel \& Hamilton
1992), while Ly$\alpha$ is a broad absorption feature in CB58 (the
lensed galaxy behind the cluster MS\,1512\,+36, Yee {\em et al.\ }1996)
and HDF\,4-555.1 (the hotdog--shaped galaxy in the HDF; Bunker {\em et
al.\ }1998).

A trend is also visible in the growth of the stellar, stellar--wind, and
ISM absorptions as one proceeds down the Fig.~1 spectra from HDF\,FF~D16
to CB58. The strength of the metallic ISM absorptions correlates with
Ly$\alpha$ absorption and anti--correlates with Ly$\alpha$ emission,
though the general shape of the local UV continuum over $\lambda\lambda
1240 - 1600$\,\AA\ suggests that the galaxies considered here all have
an ample supply of ionizing photons.  A simple scenario with a smooth
distribution of cold, neutral gas and a moderate or increasing metal
abundance level as one goes down Fig.~1 toward CB58 would be consistent
with our spectra.

The spectra of some distant star--forming galaxies, such as HDF\,FF~D16
and the mean HDF $U$--dropout spectrum of Lowenthal {\em et al.\ }(1997)
show weak P--Cygni and ISM lines suggestive of low metal abundance,
perhaps even below that of the O--stars in the SMC (near 1/5$^{\rm th}$
solar from UV spectra; Walborn {\em et al.\ }1995, Haser {\em et al.\
}1998).  Stellar winds, ISM, and the few weak photospheric absorption
lines are relatively strong, however, in HDF\,4-555.1 and CB58, implying
an abundance level near solar for these two galaxies at $z\sim 2.8$,
though abundances estimated from saturated lines should of course be
treated with caution.

The presence of weak but definite P--Cygni line profiles of the nominal
ISM absorptions from low--ionization species such as
O{\scriptsize~I}\,$\lambda$\,1302\,\AA\ and
C{\scriptsize~II}\,$\lambda$\,1335\,\AA\ is surprising: O--stars in the
Galaxy and the Magellanic clouds do not show P--Cygni profiles for their
low--ionization resonance lines, nor are these features seen in the few
high--quality HST spectra of nearby star--burst galaxies ({\em c.f.\
}Heckman \& Leitherer 1997).  One speculative idea we may offer is 
that these lines may be caused by the
onset of a ``galactic superwind'' in young systems with extreme
star--formation rates, {\em i.e.}, $> 10\,M_\odot\,{\rm yr}^{-1}$ (for a
low--density open Universe and $H_{0}\sim 50\,{\rm km\,s}^{-1}\,{\rm
Mpc}^{-1}$). Here the ionization level of outflowing gas may be lower
than in individual Galactic O--stars.  HST UV spectra of nearby
star--bursts like M82 might be helpful future comparison objects.

Dust reddening is a poorly--constrained but important issue for these
distant star--burst systems; a slightly--reddened UV continuum can lead
to substantially underestimating the true integrated luminosities.  We
have compared the spectrum of the hotdog galaxy (HDF\,4-555.1) to
IUE--based stellar models by Leitherer {\em et al.\ }(1995).  We find
that the galaxy spectrum is well fit by the oldest of Leitherer {\em et
al.}'s continuous star formation synthetic models (a population age of at
least 9 Myrs and a stellar upper mass boundary of $80\,M_\odot$). Thus
the deep UV light from the galaxy is still dominated by OB--stars. The
shape of the continuum is consistent with a visible extinction of
$A_{V}=0.2^{m}$, implying extinction $2 - 3$ times higher at the
$\lambda 1300$\,\AA\ continuum.
These conclusions are based on a poorly
known extinction curve, and the unknown placement of the OB--stars and
the geometry of the dust leads to further uncertainty. The deepest
portion of the Ly$\alpha$ absorption profile in the hotdog galaxy is
likely to arise from the stellar photospheres augmented by additional
interstellar gas; modeling suggests $N(H) \sim 10^{20}\,{\rm cm}^{-2}$
(a border--line damped Ly$\alpha$ system), roughly consistent with the
strength of the metallic ISM lines and the above--mentioned dust
extinction. An age of several tens of Myr for the dominant stellar
population is consistent with fitting the broad--band optical/near--IR
colors to the (dust--reddened) models of Bruzual \& Charlot (1993),
although this should be treated as lower limit as it is comparable to
the dynamical time, and synchronizing star bursts across the galaxy on
time--scales less than this is probably aphysical.

On the best spectra of Q\,0000~D6 we resolve two high--velocity systems
separated by $\sim 500\,{\rm km\,s}^{-1}$; large scale galactic winds
driven by SNe may be consistent with this dynamical complexity. However,
we note that Giavalisco {\em et al.\ }(1996) suggest that Q\,0000~D6 is
dynamically--relaxed, on account of its $r^{\frac{1}{4}}$ de~Vaucouleurs
profile. This is difficult to reconcile with the multiple velocity
components we observe.

\section{A Brief Glimpse at $B$--Dropouts ($z\sim 4$ Galaxies)}

It is conceptually straightforward, but observationally intensive, to
continue the Lyman--limit imaging to yet higher redshifts.  Deep imaging
in the photometric bands of $BVRI$, or the Gunn system bands including
$gri$, can be used to select $B$--dropout candidates, implying $z \sim$
4, instead of the $U$--dropouts at $z \sim$ 3 discussed above.

Our {\em entr\'{e}e} to this subfield came from $BVRI$ images around the
distant radio galaxy 6C\,0140\,+326 ($z=4.41$; Rawlings {\em et al.\
}1996). The Keck direct images go quite faint, so in September 1997 we
located $\sim 13$ potential $B$--dropouts and observed 6 of them
spectroscopically with Keck/LRIS using a slit mask ({\em c.f.\ }Dey {\em
et al.\ }1998).  These targets range in reshift from $z=3.602$ and
$z=4.020$.  None have the redshift of the (centrally positioned) radio
galaxy! Redshifts in excess of 4.5 would have been allowed by our
photometric constraints.

Four of the six galaxies show moderate to strong Ly$\alpha$ emission.
Two show broad absorption at
Ly$\alpha$. Si{\scriptsize~II}\,$\lambda$\,1260\,\AA\ is seen in
absorption in several of the six
$B$--drops. C{\scriptsize~IV}\,$\lambda\lambda$\,1548,1551\,\AA\ with a
(noisy) P--Cygni profile is detected in 4 of the galaxies. All the
systems show some Lyman forest discontinuity at $\lambda$\,1216\,\AA ;
in four of them it is quite strong with a flux ratio $\sim 2$ across the
break.

The one consistent feature of the $B$--drop spectral continua is that
the galaxy flux is at or very close to zero below $\lambda\,912$\,\AA
. Inspection of the spectrum below Ly$\alpha$ show that all have no flux
at $\lambda_{0}< 912$\,\AA , but a detectable weak continuum near
$\lambda\,1025$\,\AA\ (the Ly$\beta$ region).  Future papers will be
more quantitative about this discontinuity.  We note, however, that
since radiation that can ionize hydrogen is undetected in any of our
$B$--drop galaxies it is implausible that ionizing radiation from young
galaxies can replace the QSO ionization at $z > 4$, even considering
that the co--moving space density of radio--loud and radio--quiet
quasars is known to decline after their $z \sim 2$ peak.

One can hope that in the future the photometric high--redshift locator
techniques can be extended to even larger redshifts.

\begin{table}
\begin{center}
\caption{UV Spectral Lines from Star--forming Galaxies at $z>3$.}
\medskip
\begin{tabular}{ll}
\hline
\hline
Emission from the Gas  		& Interstellar Absorptions \cr
\hline
Ly$\alpha$ $\lambda$1216\AA	& Ly$\beta$ $\lambda$1025\AA \\
He\II~$\lambda$1640\AA	 	& Ly$\alpha$ $\lambda$1216\AA  (damped?) \\
C\III]~$\lambda$1909\AA		& Si\II~$\lambda$1260\AA \\
				& O\I~$\lambda$1302\AA \\
				& C\II~$\lambda$1335\AA \\
				& Si\IV~$\lambda\lambda$1394,1403\AA \\
				& Si\II~$\lambda$1526\AA \\
				& C\IV~$\lambda\lambda$1549\AA \\
				& Fe\II~$\lambda$1608\AA \\
				& Al\II~$\lambda$1670\AA \\
& \\
& \\
\hline
\hline
P--Cygni Wind Lines 		& Photospheric Lines \cr
\hline
N\V~$\lambda$1240\AA 	 	& C\III~$\lambda$1175\AA \\
Si\II~$\lambda$1260\AA 	 	& Si\III~$\lambda$1417\AA \\
O\I~$\lambda$1302\AA 	 	& C\III~$\lambda$1427\AA \\
C\II~$\lambda$1335\AA 	 	& S\V~$\lambda$1502\AA \\
Si\IV~$\lambda\lambda$1394,1403\AA & \\
Si\II~$\lambda$1526\AA		& \\
C\IV~$\lambda\lambda$1549\AA & \\
\hline
\end{tabular}
\end{center}
\end{table}

\newpage

\begin{acknow}
Many of the observations described here were obtained with the W.M.~Keck
Telescope. The spectrum of CB58 was kindly provided by Charles Steidel
and Max Pettini in advance of publication. HS acknowledges support by
the U.S.\ National Science Foundation through grant AST~95--28536. We
thank the Knaw meeting organizers for their hospitality.
\end{acknow}

\end{document}